\documentclass[doublecol,linenumbers]{epl2} 

\newcommand{\cont}{{\rm c}}

\newcommand{\wat}{{\rm w}}
\newcommand{\oil}{{\rm o}}
\newcommand{\ins}{{\rm ins}}
\newcommand{\sur}{{\rm sur}}
\newcommand{\ele}{{\rm ele}}
\newcommand{\os}{{\rm os}}
\newcommand{\ws}{{\rm ws}}

\newcommand{\thaoil}{{\bar{\theta}_{\rm o}}} 
\newcommand{\thoil}{{\theta_{\rm o}}}       

\title{Contact angle saturation in electrowetting: Injection of ions into the surrounding media}
\shorttitle{Ion injection mechanism of contact angle saturation}

\author{Tetsuya Yamamoto\inst{1,2}, Masao Doi\inst{1} \and David Andelman\inst{3}\thanks{E-mail: \email{andelman@post.tau.ac.il} (corresponding author)}}
\shortauthor{Yamamoto, Doi and Andelman}
\institute{
  \inst{1}Center of Soft Matter Physics and its Applications, Beihang University - Xueyuan Road 37, Beijing 100191, China\\
  \inst{2}National Composite Center, Nagoya University - Furocho, Chikusa-ku, Nagoya, 464-8603, Japan.\\
 \inst{3}School of Physics and Astronomy, Tel Aviv University - Ramat Aviv 69978, Tel Aviv, Israel \\
}


\date{October 2, 2015}

\abstract{We use the Poisson-Boltzmann theory to predict contact angle saturation of aqueous droplets in electrowetting. Our theory predicts that injection of ions from the droplet into its surrounding medium is responsible for the deviation of the apparent contact angle from the Young-Lippmann equation for large applied voltages. The ion injection substantially decreases the Maxwell stress and increases the osmotic pressure at the interface between the two media, leading to saturation of the apparent contact angle. Moreover, we find that the contact angle does not saturate, but only has a broad minimum that increases again upon further increase of the applied voltage, in agreement with experiments.}

\begin{document}

\maketitle


\section{Introduction}

Electrowetting is a process where the wettability of aqueous droplets on surfaces is controlled by an applied voltage~\cite{DavidSM2008,MugeleJPCM2005}, and is used in applications such as variable focal lens~\cite{BergeEPJE2000}, display
devices~\cite{HayesNature2003}, and lab-on-chip devices~\cite{FairLOC2002}.

In a typical experimental setup~\cite{BergePolymer1996,BergeEPJE1999}, a droplet of an aqueous (electrolyte) solution is deposited on a thin dielectric film that coats a metal working electrode (see fig.~\ref{fi:EWODgeometry}), and the space surrounding the droplet is filled with an immiscible fluid (such as oil). When a voltage is applied between a counter-electrode that is inserted into the droplet and the metal working electrode, charges carried by ions are accumulated at the interface between the droplet and the surrounding medium. The resulting Maxwell stress arising from these accumulated charges deforms the shape of the oil-water interface near the three-phase contact line. In typical experiments, the deformed region of the interface is much smaller than the droplet size, and the deformation is observed as a decrease~\cite{JonesJMM2005,KangLangmuir2002,JonesLangmuir2002} of the droplet {\it apparent} contact angle, $\bar{\theta}_\wat$, although the {\it intrinsic} contact angle, $\theta_\wat$, remains constant~\cite{MugeleSM2009,MugelePRL2003,MugeleJPCM2007}.

For small applied voltages, the apparent contact angle, $\bar{\theta}_\wat$, decreases with increasing applied voltages, following the classical Young-Lippmann equation [see eq.~(\ref{eq:EWequation})]. As the applied voltage becomes larger, $\bar{\theta}_\wat$ starts to deviate from the Young-Lippmann equation~\cite{BergeEPJE1999,MugeleSM2009,ChevalliotJAST2012} and eventually saturates --- a phenomenon called {\em contact angle saturation} (CAS). We note that in some experiments, $\bar{\theta}_\wat$ does not fully saturate but rather shows a broad minimum, beyond which it starts increasing again upon further increase of the applied voltage~\cite{ChevalliotJAST2012}.

A number of models in the past two decades have been proposed to elucidate the physical mechanism involved in CAS~\cite{KangLangmuir2002,DavidLangmuir2011,AliBiomicrofluidics2015,PapathanasiouAPL2005,PapathanasiouAPL2008,PapathanasiouLangmuir2009}. Macroscopic electrostatic considerations indicate that very large electric fields are generated near the three-phase contact line
(about ten times or larger than bulk values were reported in Ref.~\cite{PapathanasiouAPL2005}), where the oil-water interface intersects the substrate at a finite angle. Along these lines, Papathanasiou and coworkers~\cite{PapathanasiouAPL2005,PapathanasiouLangmuir2009} predicted that charges are injected and trapped inside the dielectric film, and relate it to local dielectric breakdown due to large electric fields near the contact line. The trapped charges decrease the Maxwell stress that deforms the oil-water interface, and eventually leads to saturation of the droplet contact angle. This mechanism may operate in some experimental conditions, but, in principle, one can suppress local dielectric breakdown by using alternating voltages (AC) and/or dielectric films of high dielectric strength~\cite{ChevalliotJAST2012,BergePolymer1996}. Since CAS has been observed almost universally in electrowetting experiments, saturation should be driven by a generic electrostatic mechanism that does not depend on specific experimental setups or materials. Moreover, many of the existing theories do not account for the fact that  the apparent contact angle exhibits, in some cases, a minimum (as function of the applied voltage), instead of saturation.

When the medium surrounding the droplet is a gas phase, experiments have shown an ionization of gas molecules at the contact line for voltages corresponding to the onset of CAS~\cite{BergeEPJE1999}. Moreover, molecular dynamics (MD) simulations have shown that when the total amount of charges is fixed~\cite{RobbinsPRL2012}, the apparent contact angle of nanoscopic droplets starts to deviate from the Young-Lippmann equation when charges are injected from the droplet into its surrounding medium.

Motivated by these results, we use the Poisson-Boltzmann theory to predict a physical mechanism that drives CAS. Our theory treats the statistical mechanics involved in the ion injection and thus is very different from theories that are solely based on macroscopic electrostatics~\cite{KangLangmuir2002,DavidLangmuir2011,AliBiomicrofluidics2015,PapathanasiouAPL2005,PapathanasiouAPL2008,PapathanasiouLangmuir2009}. The theory relates the phenomenon of CAS to the decrease of the Maxwell stress at the oil-water interface due to the increase in injected ions. We find that for large applied voltages, the apparent contact angle does not fully saturate. Furthermore, because of the osmotic pressure of the injected ions, the contact angle starts to increase again with increasing applied voltages, in agreement with previous experiments~\cite{ChevalliotJAST2012}. Finally, we note that our model differs from the work of Monroe {\it et al}~\cite{KornyshevPRL2006}, where the Poisson-Boltzmann theory was used to predict CAS, but in a very different system where ions do not exchange across the interface between the droplet and its surroundings.

\section{Model}

\begin{figure*}
\centering
\includegraphics[width = 0.6 \linewidth]{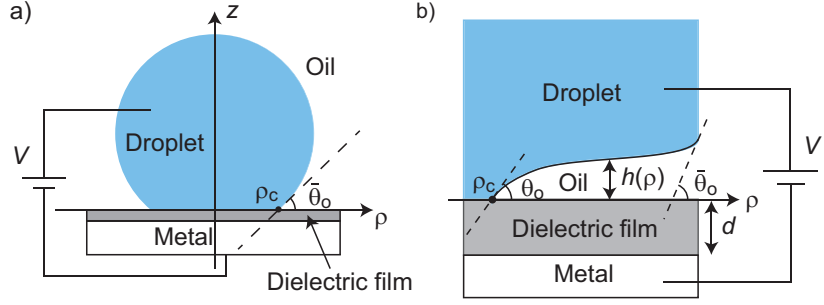} 
\caption{(a) A typical experimental setup of electrowetting. A droplet of an aqueous (electrolyte) solution is deposited on a thin dielectric film of thickness $d$, which coats a planar metal electrode. The space that surrounds the droplet is filled with oil. A voltage $V$ is applied between the bulk of droplet and the metal working electrode. Cylindrical coordinates $(\rho,\phi,z)$ are used because of the axial symmetry of the droplet. (b) An enlarged view of the oil-water interface close to the three-phase contact line, $\rho = \rho_\cont$. The shape of the oil-water interface is represented by the height function $h(\rho)$. The intrinsic contact angle between the oil-water and the planar oil-insulator interfaces is $\thoil=\pi-{\theta}_\wat$, while the apparent (macroscopically measured) contact angle is $\thaoil=\pi-\bar{\theta}_\wat$.}
\label{fi:EWODgeometry}
\end{figure*}

We take into account the injection of ions from the droplet into the surrounding medium in an extension of the electromechanical theory of electrowetting~\cite{JonesJMM2005,JonesLangmuir2002}. For convenience, we introduce the supplementary intrinsic contact angle, $\thoil = \pi - \theta_\wat$, and the supplementary apparent contact angle, $\thaoil = \pi - \bar{\theta}_\wat$. These are the angles that the oil-water interface makes with the dielectric film from the oil side (see fig.~\ref{fi:EWODgeometry}). We consider a droplet of an aqueous solution, placed on a thin dielectric film that coats a planar metal electrode, see fig. \ref{fi:EWODgeometry}. We treat cases, in which the space that surrounds the droplet is filled with oil, but our theory is applicable to cases where the surrounding medium is another immiscible fluid.

The aqueous droplet is rather large and acts as a reservoir of monovalent ions of concentration $n_\wat$. The ions in the droplet are in thermodynamic equilibrium with ions residing in the surrounding oil phase. Without an applied voltage, the ionic concentration $n_\oil$ in the oil phase has the form $n_\oil =  n_\wat {\exp}[(\mu_\wat - \mu_\oil)/k_B T]$, where $\mu_\wat$ and $\mu_\oil$ are the chemical potentials of ions in the droplet and oil phase, respectively, $k_B$ is the Boltzmann constant and $T$ is the absolute temperature. Since ions are highly insoluble in the oil phase, $\mu_\oil \gg \mu_\wat$, and the ionic concentration is exponentially small in the oil phase, $n_\oil\ll n_\wat$. Finally, as our theory treats droplet deformations on length scales much smaller than the capillary length, gravity plays no role.

The shape of the oil-water interface, which is axially symmetric, is represented by the positional vector that has the form ${\bf r}(\rho,\phi) = (\rho \cos \phi, \rho \sin \phi, h(\rho))$, where $h(\rho)$ is the interface height (along the $z$-direction) measured from the oil-film interface, see fig.~\ref{fi:EWODgeometry}. This vector describes the deformation of the oil-water interface in the proximity region of the contact line (much smaller than the droplet size). The surface of the dielectric film is hydrophobic and the contact angle, $\thoil \ll 1$, is very small. Thus, the gradient $|h'(\rho)|\ll 1$ and, to first approximation, the electric field is normal with respect to the dielectric film.

With these assumptions, the free energy of the droplet has the form
\begin{eqnarray}
F =   \int_{\rm S_\oil}  f_\sur \, {\rm d}S ~+~ \int_{\rm S} f_\ele \,{\rm d}S
~+~ \Delta P \int_{\Omega}{\rm d} V.	\label{eq:f}
\end{eqnarray}
The first term is the surface free-energy, the second term is the free-energy contribution due to the electrostatics and entropy of ions, and the third term ensures that the volume $\Omega$ of the aqueous droplet remains constant, by using a Lagrange multiplier, $\Delta P$, that is the pressure difference between the droplet and oil phase. The area integral in the first term of eq.~(\ref{eq:f}) is limited to planar section, $\rho > \rho_\cont$, covered by the oil phase ($S_\oil$), while the area integral in the second term should be performed over the entire surface ($S$) of the dielectric film (the $z = 0$ plane), and $\Omega$ is the system volume.

The surface free-energy, $f_\sur$, has the form
\begin{eqnarray}
f_\sur(\rho) \simeq \gamma_{\os} - \gamma_{\ws} + \gamma \left( 1 + \frac{1}{2} {h'}^2(\rho) \right),
\end{eqnarray}
where $\gamma_\os$, $\gamma_\ws$, and $\gamma$ are the surface tensions of the oil-substrate, water-substrate, and oil-water interfaces, respectively.
We take the surface free-energy of the droplet at zero voltage as our reference state.
Retaining only the leading order term, ${\sim}(h')^2$, in the gradient expansion, $|h'(\rho)|$

The free-energy contribution, $f_\ele$, has the form
\begin{eqnarray}
& &\frac{e^2 f_\ele(\rho)}{(k_B T)^2} = \nonumber \\
 & &- \epsilon_\oil \int_0^{h} dz \left[ \kappa_\oil^2 (\cosh \psi_\oil - 1) +
\frac{1}{2} \left( \frac{\partial \psi_\oil}{\partial z} \right)^2 \right] \nonumber \\
& & - \epsilon_\wat \int_h^\infty dz \left[ \kappa_\wat^2 (\cosh \psi_\wat - 1) + \frac{1}{2}
\left( \frac{\partial \psi_\wat}{\partial z} \right)^2 \right] \nonumber \\
& & - \frac{1}{2} \epsilon_\ins \int_{-d}^0 dz \left( \frac{\partial \psi_\ins}{\partial z}
\right)^2 - ~ q_\ele(\rho) U\, , \label{eq:fele}
\end{eqnarray}
where $\psi_\wat(\rho,z)$, $\psi_\oil(\rho,z)$, and $\psi_\ins(\rho,z)$ are the dimensionless local electrostatic potentials (rescaled by $e/k_B T$, where $e$ is the elementary charge). The subscripts `$\wat$', `$\oil$', and `$\ins$' indicate the aqueous droplet ($z > h(\rho)$), the oil phase ($0 < z < h(\rho)$), and the dielectric film ($- d < z < 0$), respectively, with the corresponding dielectric constants: $\epsilon_\wat$, $\epsilon_\oil$, and $\epsilon_\ins$. The inverse Debye length in the droplet and oil phase is, respectively, $\kappa_\wat \equiv \sqrt{8 \pi l_\wat n_\wat}$ and $\kappa_\oil \equiv \sqrt{8 \pi l_\oil n_\oil}$, where $l_\wat \equiv e^2/(4 \pi \epsilon_\wat k_B T)$ and $l_\oil \equiv e^2/(4 \pi \epsilon_\oil k_B T)$ are the Bjerrum lengths in the corresponding regions. Finally, in the last term $q_\ele(\rho)~= - \epsilon_\ins \frac{\partial}{\partial z} \psi_\ins(\rho,z) \mid_{z = - d}$ is the charge density (rescaled by $k_BT/e$) on the electrode at $z=-d$, and $U \equiv eV/(k_B T)$ is the dimensionless potential to be used hereafter, proportional to $V$, the voltage applied between the droplet and the metal working electrode (see fig.~\ref{fi:EWODgeometry}).

Minimizing the free energy, eq.~(\ref{eq:f}), with respect to the three electrostatic potentials leads to a set of Poisson-Boltzmann equations:
\begin{eqnarray}
\frac{\partial^2 \psi_\wat} {\partial z^2}&=& \kappa_\wat^2 \sinh \psi_\wat(\rho,z),	\label{eq:PBequation/wat} \\
\frac{\partial^2 \psi_\oil}{\partial z^2} &=& \kappa_\oil^2 \sinh \psi_\oil(\rho,z), 	\label{eq:PBequation/oil} \\
\frac{\partial^2 \psi_\ins}{\partial z^2} &=& 0.	\label{eq:PBequation/ins}
\end{eqnarray}
We treat the dielectric film as a perfect insulator, where ions cannot penetrate. This is in contrast to the case treated in refs.~\cite{PapathanasiouAPL2005,PapathanasiouLangmuir2009}. Equations~(\ref{eq:PBequation/wat})-(\ref{eq:PBequation/ins}) should be solved with the following boundary conditions: i) the electrostatic potential is zero in the bulk of the droplet, ii) the electrostatic potential is $- U$ on the $z=-d$ electrode, and iii) the electrostatic potential and electric displacement vector (in the $z$-direction) are continuous at the oil-water interface, $z = h(\rho)$, and at the oil-insulator interface, $z = 0$. Equations~(\ref{eq:PBequation/wat}) and (\ref{eq:PBequation/oil}), as well as the boundary conditions at the oil-water interface, ensure the continuity of electrochemical potential at the oil-water interface for both cations and anions; our theory takes into account explicitly the scenario that ions can be injected from the droplet into the oil phase and vice versa, due to the applied voltage.

Minimizing the free energy, eq.~(\ref{eq:f}), with respect to the position $h(\rho)$ of the oil-water interface leads to a force balance equation of the form
\begin{eqnarray}
\Delta P - \gamma \frac{1}{\rho} \frac{d}{d \rho} \left( \rho \frac{d}{d \rho} h(\rho) \right)
- 2 n_\oil k_B T (\widehat{\Pi}_\oil(h) - 1) = 0,
\label{eq:forcebalance}
\end{eqnarray}
where $\Delta P$  of the first term is the pressure difference, the second term is the capillary  force, and the third one
is related to the electro-osmotic pressure~\cite{SamBook4_2}. The dimensionless electro-osmotic pressure, $\widehat{\Pi}_\oil(h)$, in the oil phase has the form
\begin{eqnarray}
\widehat{\Pi}_\oil(h) = \cosh \psi_\oil(\rho,z) - \frac{1}{2 \kappa_\oil^2}
\left( \frac{\partial\psi_\oil}{\partial z}  \right)^2,	\label{eq:Pio}
\end{eqnarray}
where its first term accounts for the osmotic pressure of the ions and the second one for the electrostatic Maxwell stress. It can be shown that $\widehat{\Pi}_\oil(h)$ is equal to the integration constant of the first integral of eq.~(\ref{eq:PBequation/oil}). Thus, it depends on $\rho$ only via the position $h(\rho)$ of the oil-water interface. Note that the electro-osmotic pressure in the aqueous droplet does not contribute to the force balance in eq.~(\ref{eq:forcebalance}), because its value at the oil-water interface is equal to its bulk droplet value, $2 n_\wat k_B T$. We emphasize that the osmotic pressure of the ions --- the first term of eq.~(\ref{eq:Pio}) --- is a new ingredient that has not been previously considered in works that employed macroscopic electrostatics.

The boundary condition used to solve eq.~(\ref{eq:forcebalance}) is that the oil-water interface intersects the surface of the dielectric film at the contact line, $h(\rho_\cont) = 0$, with an angle $\thoil=h'(\rho_\cont)$. Finally, by minimizing eq.~(\ref{eq:f}) with respect to the position $\rho = \rho_\cont$ of the contact line,
we obtain that the intrinsic contact angle, $\thoil \simeq \sqrt{2 (\gamma_\os - \gamma_\ws + \gamma)/\gamma}$~\cite{SamBook4_2}. Our theory thus predicts that the angle $\thoil$ does not depend on the applied voltage, in agreement with refs.~\cite{MugeleSM2009,MugelePRL2003,MugeleJPCM2007}.

Without loss of generality, we assume that the curvature $h'(\rho)/\rho$ of the oil-water interface in the radial direction is quite small [see also the second term of eq.~(\ref{eq:forcebalance})]. The first integral of eq.~(\ref{eq:forcebalance}) is obtained by multiplying both sides of this equation by $h'(\rho)$, and then integrating it with respect to $\rho$ with the boundary condition, $h'(\rho_\cont) = \thoil$. Note that the oil-water interface is deformed by the applied electric field in a region that is much smaller than the radius $\gamma/(2 \Delta P)$ of the droplet, $h \ll \gamma/(2 \Delta P)$~\cite{MugeleSM2009,MugelePRL2003,MugeleJPCM2007}. In length scales comparable to the droplet radius, the deformation of the oil-water interface is observed as a decrease of the contact angle, and the limit $\thaoil \equiv \lim_{\rho \to \infty} h'(\rho)$, is defined as  the {\it apparent} contact angle. Then, first integral of eq.~(\ref{eq:forcebalance}) yields the form
\begin{eqnarray}
\cos \thaoil = \cos \thoil + \frac{2 n_\oil k_B T}{\gamma} \int_0^\infty {\rm d} h
\left[\widehat{\Pi}_\oil(h) - 1\right] \, ,	\label{eq:CAequation/general}
\end{eqnarray}
where $\cos x \simeq 1 - x^2/2$ is used for both the intrinsic and apparent
contact angles, $\thoil$ and $\thaoil$.

The electro-osmotic pressure, $\widehat{\Pi}_\oil(h)$ in eq.~(\ref{eq:CAequation/general}), is derived by using eq.~(\ref{eq:Pio}), where the electrostatic potential $\psi_\oil(\rho,z)$ is obtained from eqs.~(\ref{eq:PBequation/wat})--(\ref{eq:PBequation/ins}) with the corresponding boundary conditions (shown below these equations). Considering the high salt limit inside the droplet, the ions strongly screen the electric field and the use of  Debye-H\"uckel approximation, $\sinh \psi_\wat \simeq \psi_\wat$ in eq.~(\ref{eq:PBequation/wat}), is justified. The electrostatic potential and electric field in the droplet thus have the approximate form
\begin{eqnarray}
\psi_\wat(\rho,z) &=& \psi_\wat(h) {\rm e}^{- \kappa_\wat (z - h)} \nonumber\\
E_\wat(\rho,z) &=& \kappa_\wat \psi_\wat(h) {\rm e}^{- \kappa_\wat (z - h)},
\end{eqnarray}
where $\psi_\wat$ and $E_\wat=-\partial \psi_\wat/\partial z$ depend on the radial coordinate $\rho$ only via the height $h(\rho)$ of the oil-water interface.

In contrast, the ionic concentration in the oil phase is very small, unless ions are injected from the droplet into the oil phase by the applied voltage. With positive applied voltages, $U > 0$, cations are injected from the droplet into the oil phase, and likewise, anions are injected from the oil phase into the droplet. When the applied voltage is large enough, most of the ions in the oil phase are cations, $\sinh \psi_\oil \simeq - \frac{1}{2} {\exp}(- \psi_\oil)$. In this case, the Poisson-Boltzmann equation in the oil phase, eq.~(\ref{eq:PBequation/oil}), is non-linear and its non-linearity plays a crucial role in driving CAS.

\section{Results}

\begin{figure}
\centering
\includegraphics[width = 0.7 \linewidth]{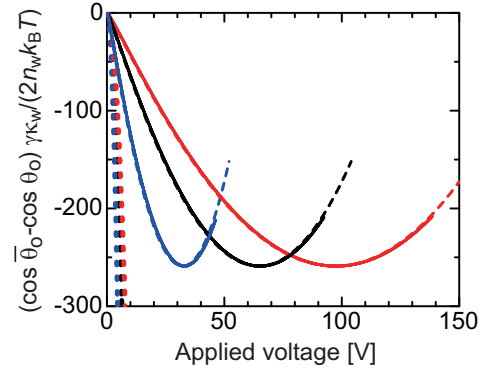} 
\caption{The difference $\cos\thaoil - \cos\thoil$ of apparent and intrinsic contact angles (rescaled by $2 n_\wat k_B T/\gamma \kappa_\wat$) is plotted as function of the applied voltage $V$ (in volts) (solid lines). The rescaled thickness of the dielectric films $d(\epsilon_\oil \kappa_\oil /\epsilon_\ins)$ is $0.5 \times 10^{-2}$ (blue), $1.0 \times 10^{-2}$ (black), and $1.5 \times 10^{-2}$ (red). The predictions of the Young-Lippmann equation (\ref{eq:EWequation}) are shown for comparison as dotted lines, and the asymptotic predictions of eq.~(\ref{eq:SATequation}) for large $V$ are shown as dashed lines. The rescaling factor $2 n_\wat k_B T/\gamma \kappa_\wat$ is about $6.8 \times 10^{-3}$ for ionic concentration  of $0.1\,{\rm M}$ in the droplet, and the ratio $\epsilon_\oil \kappa_\oil/\epsilon_\wat \kappa_\wat$ is fixed to $1.0 \times 10^{-4}$.}
\label{fi:contactangle}
\end{figure}

\begin{figure}
\centering
\includegraphics[width = 0.75 \linewidth]{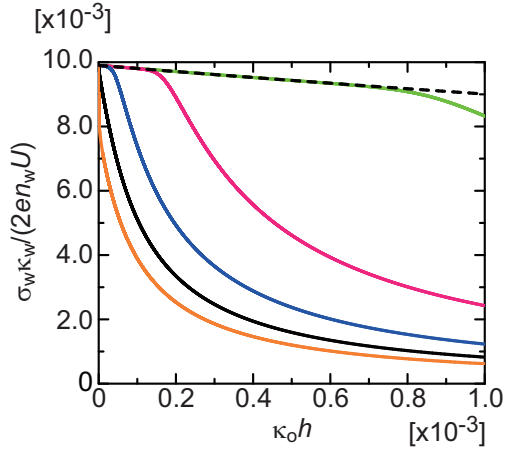} 
\caption{Rescaled local capacitance $\sigma_\wat \kappa_\wat/(2 e n_\wat U)$  plotted as function of the rescaled height $\kappa_\oil h$ of the oil-water interface for several values of the applied voltage: $V=5\,{\rm [V]}$ (light green), $20\,{\rm [V]}$ (magenta), $40\,{\rm [V]}$ (blue), $60\,{\rm [V]}$ (black), and $80\,{\rm [V]}$ (orange). The local capacitance $\sigma_\wat/U$ is defined by the charge density $\sigma_\wat$  (per unit area) at the oil-water interface divided by the applied voltage, $U$. Other parameters are $\epsilon_\oil \kappa_\oil d/\epsilon_\ins = 0.01$ and $\epsilon_\oil \kappa_\oil/\epsilon_\wat \kappa_\wat = 1.0 \times 10^{-4}$. For comparison, the (black) dashed line shows the local capacitance $(h/\epsilon_\oil+d/\epsilon_\ins)^{-1}$, rescaled as well by $2 e n_\wat/\kappa_\wat$. It corresponds to the case where there are no injected ions into the oil phase.}
\label{fi:localcapacitance}
\end{figure}

We calculate the angle $\bar{\theta}_\oil(V)$ that is supplementary to the apparent contact angle, $\bar{\theta}_{\rm w}$, as function of the applied voltage, $V$, by using eq.~(\ref{eq:CAequation/general}) (see the solid lines in fig.~\ref{fi:contactangle}).
By changing the integration variable in eq.~(\ref{eq:CAequation/general}) from $h$ to the electro-osmotic
pressure, $\widehat{\Pi}_\oil$ (employing the fact that $h = 0$ at one boundary and
$\widehat{\Pi}_\oil = 1$ at the other boundary) and expressing the height $h$ as a
function of $\widehat{\Pi}_\oil$ from eq.~(\ref{eq:Pio}), one can perform analytically the integration in eq.~(\ref{eq:CAequation/general}).
For small $V$, eq.~(\ref{eq:CAequation/general}) reduces to the classical Young-Lippmann form
\begin{eqnarray}
\cos \thaoil \simeq \cos \thoil - \frac{1}{2} \frac{\epsilon_\ins}{\gamma d} V^2,	\label{eq:EWequation}
\end{eqnarray}
where $V$ is the applied voltage in volts (see the dotted lines in fig.~\ref{fi:contactangle}). We get this match because the applied voltage is not large enough to drive the ion injection from the droplet into the oil phase. As a further check, we computed directly the combined capacitance of the oil phase and dielectric film, $(h/\epsilon_\oil + d/\epsilon_\ins)^{-1}$. As anticipated this capacitance matches our general expression in the low-$V$ limit (see the dashed and light green lines in fig.~\ref{fi:localcapacitance}), as it accounts for the charges that accumulate at the oil-water interface when ion injection is not allowed.

For larger (and positive) values of $U=eV/k_B T$, $\thaoil$ starts to deviate from the Young-Lippmann equation~(\ref{eq:EWequation}) (fig.~\ref{fi:contactangle}). Cations are injected from the droplet into the oil phase and change the electric field  (fig.~\ref{fi:localcapacitance}). The function of $\cos \thaoil$ shows a broad peak at a threshold voltage $U^*$, which depends on the (rescaled) thickness $d(\epsilon_\oil \kappa_\oil /\epsilon_\ins)$ of the dielectric film. For $U > U^*$, $\cos \thaoil$ increases slowly. When the peak of $\cos \thaoil$ is broad enough, it can be experimentally observed as a `saturation'. Hence, our model predicts that CAS is not a real saturation of $\thaoil$, but rather a broad minimum as function of the applied voltage. This finding is in agreement with recent experiments that show that the apparent contact angle starts to increase again for $U > U^*$~\cite{ChevalliotJAST2012}.

Another result obtained for the large $U$ limit is an asymptotic form of $\cos \thaoil$
\begin{eqnarray}
& &\cos \thaoil -
  \cos \thoil \simeq \nonumber\\
  & &\nonumber\\
  && \frac{2 n_\oil k_B T}{\kappa_\oil \gamma}
\left[ - \frac{2\epsilon_\ins U}{\epsilon_\oil \kappa_\oil d} \left \{ \ln \left( \frac{\epsilon_\ins U}
{\epsilon_\oil \kappa_\oil d} \right) - 1 \right \} \right. \nonumber \\
& & \left. + \frac{1}{2} \frac{\epsilon_\oil \kappa_\oil}{\epsilon_\wat \kappa_\wat} \left( \frac{\epsilon_\ins U}
{\epsilon_\oil \kappa_\oil d} \right)^2  \right],	\label{eq:SATequation}
\end{eqnarray}
where the limit $\epsilon_\ins/d \ll \epsilon_\wat \kappa_\wat$ is used (see the dashed lines in fig.~\ref{fi:contactangle}). Note that the parameter $\epsilon_\oil \kappa_\oil/(\epsilon_\wat \kappa_\wat) = \sqrt{{\epsilon_\oil}/{\epsilon_\wat}} \exp[(\mu_\wat - \mu_\oil)/(2 k_B T)]$ depends only on the specific material parameters of added salt and oil. The asymptotic expression, eq.~(\ref{eq:SATequation}), predicts that the value of $\cos \thaoil$ at $U = U^*$ depends on $n_\oil^{1/2}\gamma^{-1}$, but not on the thickness $d$ or the film dielectric constant, $\epsilon_\ins$ (see fig.~\ref{fi:contactangle}). Moreover, it can be shown that at $U = U^*$,  $\cos \thaoil -\cos \thoil \sim - {2 n_\wat k_B T}({\gamma \kappa_\wat})^{-1} \log (\epsilon_\wat \kappa_\wat/(\kappa_\oil \epsilon_\oil))$, and $U^* \sim \epsilon_\wat \kappa_\wat d/\epsilon_\ins$ for small values of $\epsilon_\wat \kappa_\wat/(\kappa_\oil \epsilon_\oil)$.

\begin{figure}
\centering
\includegraphics[width = 0.7 \linewidth]{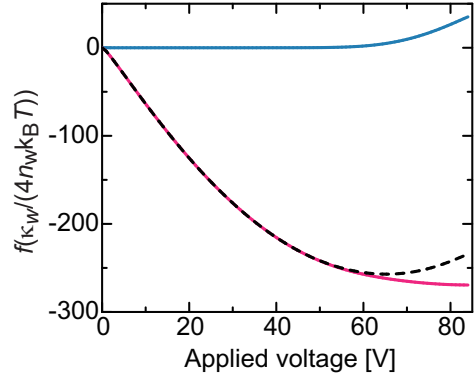} 
\caption{The two forces acting at the oil-water interface (rescaled by $4 n_\wat
k_B T/\kappa_\wat$) are shown as function of the applied voltage, $V$ (in  volts). These forces are calculated from the integral of the Maxwell stress (magenta) and osmotic pressure (cyan) along the oil-water interface (see the second term of eq. (\ref{eq:CAequation/general})). The sum of these forces is proportional to $\cos \thaoil - \cos \thoil$ (black broken). The parameters used  are: $\epsilon_\oil \kappa_\oil d/\epsilon_\ins = 0.01$ and $\epsilon_\oil \kappa_\oil/\epsilon_\wat \kappa_\wat = 1.0 \times 10^{-4}$.}
\label{fi:separation}
\end{figure}

For small applied voltages, the Maxwell stress arising from charges accumulated at the oil-water interface dominates over the osmotic pressure arising from the injected ions (see the magenta and cyan lines in fig.~\ref{fi:separation}). The charge density at the interface decreases with the increasing of injected charges, and decreases the Maxwell stress there. The deviation of the Maxwell stress from the classical Young-Lippmann equation increases with increasing $U$, and eventually, the Maxwell stress saturates (see the magenta line in fig.~\ref{fi:separation}). The difference, $\cos \thaoil - \cos \thoil$, is proportional to the electro-osmotic pressure applied to the oil-water interface [see eq.~(\ref{eq:CAequation/general})]; without the osmotic pressure of the injected ions, the apparent contact angle indeed saturates.

\section{Discussion}

We find that injection of ions from an aqueous droplet into its surrounding oil phase offers a physical mechanism that drives contact angle saturation (CAS). The Maxwell stress at the oil-water interface decreases with increasing injected charges. Furthermore, the apparent contact angle, $\bar{\theta}_\oil$, of the droplet does not saturate, but rather shows a broad minimum, followed by an increase of $\bar{\theta}_\oil$ for larger applied voltages, in agreement with experiments~\cite{ChevalliotJAST2012}. This behavior of $\thaoil$ is driven by the osmotic pressure generated by the injected ions.

Our theory treats the dielectric film, which is used to insulate the droplet from the substrate electrode, as a perfect insulator. This has to be compared with the  theories of Papathanasiou and coworkers~\cite{PapathanasiouAPL2005,PapathanasiouAPL2008,PapathanasiouLangmuir2009}, which predicted that CAS is driven by a local dielectric breakdown of the dielectric film. The breakdown mechanism may apply in experiments that use dielectric films of relatively small dielectric strength. When CAS is driven by a such a dielectric breakdown, the apparent contact angle, $\bar{\theta}_\oil$, shows a hysteresis because dielectric breakdown is an irreversible process. Furthermore, $\bar{\theta}_\oil$ will not show an increase for applied voltages that are larger than the CAS threshold $U^*$. On the other hand, the mechanism proposed in this Letter is generic and valid even for dielectric films of larger dielectric strength. In addition, our predicted $\bar{\theta}_\oil$ is reversible and shows a minimum, beyond which its value increases with the applied voltage.

An important underlying assumption is that the droplet and the surrounding medium are in thermodynamic equilibrium. However, this may not be the case in some experiments because of the time scale associated with ion injection from the droplet into the surrounding medium and the further diffusion in the latter medium. In particular, the slower ion dynamics may be important when the droplet is surrounded by a gas phase where molecules are very dilute. We note that in some experiments~\cite{ChevalliotJAST2012}, the apparent contact angle is not very sensitive to the ionic concentration in the droplet. This may be understood in terms of a non-equilibrated situation for which our theory would not be applicable.

Experiments by Chevalliot and coworkers~\cite{ChevalliotJAST2012} suggest that for AC voltages with moderate frequency, CAS is driven by a process that is faster than the local dielectric breakdown, where this fast process may be driven by the injection of ions from the droplet into the oil phase. In the future, it will be of interest to capture the ionic motion, {\it e.g.,} by impedance spectroscopy that measures capacitive (displacement) currents between the droplet and the metal working electrode. Such experiments can be complemented by extending our theory to systems where electrowetting is driven by applied AC voltages.

For small applied voltages, the angle $\thaoil$ reduces to the Young-Lippmann equation, as in eq.~(\ref{eq:EWequation}), whereas for large applied voltages,
$\thaoil$ has an asymptotic form, as in eq.~(\ref{eq:SATequation}). Equation~(\ref{eq:SATequation}) also predicts that $\kappa_\wat \gamma(\cos \thaoil - \cos \thoil)/(2 n_\wat k_B T)$ scales with the rescaled applied voltage, $\epsilon_\ins U/(\epsilon_\oil \kappa_\oil d)$. This implies that plots of different experimental conditions should collapse into one curve, provided that the specific combination of material parameters of the ions and oil, as in $\epsilon_\oil \kappa_\oil/(\epsilon_\wat \kappa_\wat)$, remains fixed. Moreover, we also predict that the apparent contact angle at saturation depends on parameters that characterize the ionic solvation in the surrounding oil phase.

We hope that future experiments will test the predictions presented in the Letter, and, in general, will advance the understanding of the
principle mechanism underlying contact angle saturation in electrowetting.

\acknowledgments
This research was supported by the ISF-NSFC joint research program (grant No.\ 885/15).
DA acknowledges support
from the Israel Science Foundation (ISF) (grant No.\ 438/12) and the United
States--Israel Binational Science Foundation (BSF) (grant No.\ 2012/060).
MD acknowledges support from the Chinese Central Government under the One Thousand Talent Program.



\begin{thebibliography}{99}

\bibitem{DavidSM2008}
  \Name{Shamai R., Andelman D., Berge B., and Hayes R.}
  \REVIEW{Soft Matter}{4}{2008}{38}.

\bibitem{MugeleJPCM2005}
  \Name{Mugele F. and Baret J. C.}
  \REVIEW{J. Phys.: Condens. Matter}{17}{2005}{R705}.

\bibitem{BergeEPJE2000}
  \Name{Berge B. and Peseux J.}
  \REVIEW{Eur. Phys. J. E}{3}{2000}{159}.

\bibitem{HayesNature2003}
  \Name{Hayes R. A. and Feenstra B. J.}
  \REVIEW{Nature}{425}{2003}{383}.

\bibitem{FairLOC2002}
  \Name{Pollack M. G., Shenderov A. D., and Fair R. B.}
  \REVIEW{Lab Chip}{2}{2002}{96}.

\bibitem{BergePolymer1996}
  \Name{Vallet M., Berge B., and Vovelle L.}
  \REVIEW{Polymer}{37}{1996}{2465}.

\bibitem{BergeEPJE1999}
  \Name{Vallet M., Vallade M., and Berge B.}
  \REVIEW{Eur. Phys. J. B}{11}{1999}{583}.

\bibitem{JonesLangmuir2002}
  \Name{Jones T. B.}
  \REVIEW{Langmuir}{18}{2002}{4437}.

\bibitem{JonesJMM2005}
  \Name{Jones T. B.}
  \REVIEW{J. Micromech. Microeng.}{15}{2005}{1184}.

\bibitem{KangLangmuir2002}
  \Name{Kang K. H.}
  \REVIEW{Langmuir}{18}{2002}{10318}.

\bibitem{MugeleSM2009}
  \Name{Mugele F.}
  \REVIEW{Soft Matter}{5}{2009}{3377}.

\bibitem{MugelePRL2003}
  \Name{Buehrle J., Hermingaus S., and Mugele F.}
  \REVIEW{Phys. Rev. Lett.}{91}{2003}{086101}.

\bibitem{MugeleJPCM2007}
  \Name{Mugele F. and Buehrle J.}
  \REVIEW{J. Phys.: Condens. Matter}{19}{2007}{375112}.

\bibitem{ChevalliotJAST2012}
  \Name{Chevalliot S. and Kuiper S. and Heikenfeld J.}
  \REVIEW{J. Adhesion Sci. Technol.}{26}{2012}{1909}.

\bibitem{DavidLangmuir2011}
  \Name{Klarman D. and Andelman D.}
  \REVIEW{Langmuir}{27}{2011}{6031}.

\bibitem{AliBiomicrofluidics2015}
  \Name{Ali H. A. A., Mohamed H. A., and Abdelgawad M.}
  \REVIEW{Biomicrofluidics}{9}{2015}{014115}.

\bibitem{PapathanasiouAPL2005}
  \Name{Papathanasiou A. G. and Boudouvis A. G.}
  \REVIEW{Appl. Phys. Lett.}{86}{2005}{164102}.

\bibitem{PapathanasiouLangmuir2009}
  \Name{Drygiannakis A. I., Papathanasiou A. G, and Boudouvis A. G.}
  \REVIEW{Langmuir}{25}{2009}{147}.

\bibitem{PapathanasiouAPL2008}
  \Name{Papathanasiou A. G., Papaioannou, and Boudouvis A. G.}
  \REVIEW{Appl. Phys. Lett.}{103}{2008}{034901}.

\bibitem{RobbinsPRL2012}
  \Name{Liu J., Wang M., Chen S., and Robbins M. O.}
  \REVIEW{Phys. Rev. Lett.}{108}{2012}{216101}.

\bibitem{KornyshevPRL2006}
  \Name{Monroe C. W., Daikhin L. L., Urbakh M., and Kornyshev A. A.}
  \REVIEW{Phys. Rev. Lett.}{97}{2006}{136102}.


\bibitem{SamBook4_2}
For more details, see Chapter 4.2 in
\Name{Safran S. A.}
\Book{Statistical Thermodynamics of Surfaces, Interfaces, and Membranes}
\Publ{Westview Press, Boulder, CO, USA}
\Year{2003}.




\end{thebibliography}
\end{document}